\pdfoutput=1

\documentclass[aps,preprintnumbers,amsmath,amssymb,twocolumn, tightenlines,superscriptaddress,nofootinbib]{revtex4}
\usepackage{graphicx}
\usepackage{slashed}
\usepackage{lipsum}
\usepackage{tikz,mathpazo}
\usetikzlibrary{shapes.geometric, arrows}
\usepackage[none]{hyphenat}
\usepackage[english]{babel}
\begin{document}

%\twocolumn[\hsize\textwidth\columnwidth\hsize\csname @twocolumnfalse\endcsname

\title{Chiral vortical catalysis}

\author {Yin Jiang} %\email{}
\address{Physics Department, Beihang University, 37 Xueyuan Rd, Beijing 100191, China}
\date{\today}

\begin{abstract}
Gluon interaction introduces remarkable corrections to the magnetic polarization effects on the chiral fermions, which is known as the inverse magnetic catalysis. It is a natural speculation that the vorticity, which has many similar properties as magnetic field, would bring non-negligible contribution to the chiral rotational suppression.  
Using the intuitive semi-classical background field method we studied the rotation dependence of the effective strong interaction coupling. Contrary to the magnetic field case the rotation increases the effective coupling which leads to slowing down the condensate melting procedure with temperature. This could be named as the chiral vortical catalysis or inverse rotation suppression. Imposing such dependence to the coupling in the NJL model, we numerically checked this analysis qualitatively. The pseudo critical temperature is shown to rise with the rotation and approach saturation eventually which may be induced by the model cutoff.
\end{abstract}

\maketitle

{\it {Introduction}.---}
Quantum effects induced by the strong magnetic and vortical field on chiral fermions  have been widely studied in condensed matter\cite{Rylands:2021dds,berthier2001high}, cold atom\cite{Aidelsburger:2017qlh, Babik, Zhang}, astrophysics\cite{Watts:2016uzu,Grenier:2015pya} and high energy nuclear physics\cite{Chen:2015hfc,Fukushima:2018grm} recently. In laboratory several novel chiral effects have been observed in systems of cold atom and condensed matter, such as chiral magnetic effect, magnetic catalysis and chiral vortical effect. Undoubtedly these experiments require strong field generators, delicate field controllers as well as novel material samples which could excite chiral collective modes. In high energy nuclear physics studies of magnetic and vortical effects are motivated by the extremely large corresponding background fields generated in relativistic heavy ion collisions and the fundamental chiral fermions-- light flavor quarks in the produced quark-gluon plasma(QGP). As simulations of phenomenological models and measurements of hadron polarization at the Relativistic Heavy Ion Collider(RHIC) and the Large Hadron Collider(LHC) indicate, the peak magnetic field would exceed $10^{14}$T and local rotation $10^{24}$Hz in QGP\cite{Skokov:2009qp, Voronyuk:2011jd, Deng:2012pc, Becattini:2015ska,Jiang:2016woz}. And the 
remaining large amount of collision energy would heat the QGP to hotter than 165MeV that is supposed to melt chiral condensate easily and make light flavor quarks turn to approximate chiral fermions. Therefore the chiral magnetic/vortical effects are expected to take place in the QGP doubtlessly. And corresponding measurable signals would be detected in the heavy ion experiments if strong fluctuations can be dealt with systematically and deliberately\cite{Wang:2012qs,Aziz:2020nia,Haque:2018jht,Li:2020dwr}. 

Theoretically these chiral effects can be roughly divided into three classes. The first includes effects that can be understood by the polarization energy shift of chiral fermions in present of background field, such as the condensate catalysis or inhibition\cite{Fukushima:2018grm,Jiang:2016wvv}. In the context of different systems, such effects usually involve different fermion pairings, for example the fermion-antifermion pairing in the chiral magnetic catalysis, vortical inhibition and di-fermion pairing in superconductivity. The second class is usually described by the transport theory of single chiral fermion in the magnetic/vortical field\cite{Fukushima:2008xe,Kharzeev:2013ffa,Sadofyev:2010pr,Avkhadiev:2017fxj,Huang:2018aly}. They are just the semi-classical expansion of the full quantum state in certain background fields and essentially induced by the non-trivial topological properties of chiral spinor's wave function in the momentum space. This can be systematically studied with the relativistic quantum kinetic theory\cite{Weickgenannt:2019dks,Gao:2019znl}. The famous chiral magnetic/vortical effects are in this class. Because of single particle properties such effects are expected to be generic and behave almost the same in different systems, such as condensed matter, cold atom or QGP. In other words, they have little relationship with the interaction between fermions, which usually encode the most non-trivial part of a certain system. Go beyond the single fermion  and consider novel quantum corrections to interaction, we arrive at the last class. In the context of quantum chormodynamics(QCD) a typical example is the inverse magnetic catalysis(IMC) which was observed in lattice QCD simulation around a decade ago\cite{Bandyopadhyay:2020zte,Bali:2012zg}. In such simulations the chiral condensate appears to melt more easily at larger background magnetic field which should have helped the condensate to increase in the scenario of single fermion. Hence in IMC the chiral condensate behavior can not be completely understood by magnetic polarization of light flaovr fermions(quarks) any longer. The gluon part, which even has no {\it direct} interaction with the background magnetic field, will qualitatively bend the increasing curve of the critical temperature along magnetic field. Rotation and magnetic field share many of the same characteristics\cite{Chen:2015hfc,Fukushima:2018grm}. Analogy to the magnetic catalysis and chiral magnetic effect in the class 1 and 2 we have similar effects known as chiral rotational suppression/inhibition and chiral vortical effects respectively. What about the class 3? Will the gluons' corrections behave non-trivially again in the rotational case? We will find some clues to this question in this work.
%In this work we will explore the influence of vorticity on the QCD effective running coupling and its impact on the phase structure.

%{\color{red}For vorticity itself, after a inspiring qualitative agreement of the simulations of the $\Lambda$ polarization dependence on the collision energy and centrality with the STAR measurements\cite{STAR:2017ckg,Xie:2017upb}, people found that the vorticity distribution and evolution in QGP may be much more complicated by considering both the quantitative analysis on the global polarization and the qualitative mismatch of the local polarization profiles\cite{Shapoval:2017jej,Li:2017dan, Becattini:2020ngo}.}
%In order to understand the vector meson polarization measurements\cite{BedangadasMohantyfortheALICE:2017xgh,Zhou:2019lun,Mohanty:2020bqq}, in the framework of kinetic theory, quantum Wigner functions and hydrodynamics, various attempts\cite{Kharzeev:2013ffa,Li:2017dan,Karpenko:2021wdm, Avkhadiev:2017fxj, Gao:2019znl,Weickgenannt:2019dks,DelZanna:2013eua,Huang:2018aly,Huang:2020kik, Karpenko:2013wva} have been made to mend the usual simulation which is basing on the straightforward scenario of the parton collisions. In most of these works, valence quraks, which serve as spin-carriers of final-state hadrons, have attracted almost all the interests\cite{Yang:2017sdk,Sheng:2019kmk}. The gluons, which carrying double spin number and thus suffered double polarization effects, are not treated seriously enough. 

A strong background field will change the vacuum structure of a system drastically in both classical and quantum mechanics\cite{Fukushima:2018grm}, e.g. in a uniformly rotating or magnetized system the eigen stats are Bessel or confluent hypergeometric functions\cite{Chen:2015hfc}. The loss of translational invariance will make the straightforward loop computations of the vertices much more difficult. In this work we adopt the very intuitive and novel method\cite{Nielsen:1980sx} to sidestep the loop computation involving complicated summations of special function series. The idea of the approach is computing the {\it vacuum } energy of a certain system in present of semi-classical background color magnetic field by summing over all the eigen energy levels $E_{vac}\sim \sum E_n$. And the the effective coupling constant will be obtained by extracting the effective permeability $\mu$ and the corresponding dielectric constant $\epsilon=\mu^{-1}$ by comparing the vacuum energy to the standard form $E_{vac}=-4\pi\chi^2B^2/2$. This method has reproduced the famous asymptotic freedom formula successfully and been applied to study the coupling running profile at finite temperature and chemical potential\cite{Schneider:2002jc,Schneider:2002sg,Schneider:2003uz}. Introducing such qualitative dependence into a effective model, the IMC has also been reproduced successfully\cite{Farias:2014eca}. In this work we will first determine the rotation dependence of the effective coupling constant in a pure gluon system for simplicity, and then switch to a pure fermion system to study the chiral restoration at finite temperature with Nambu--Jona-Lasinio(NJL) model by introducing the same rotation dependence to the effective coupling constant. It will be shown that the rotation will increase the coupling and thus make the condensate harder to melt with temperature increasing which appears to be completely contrary to the magnetic case. 
%The vorticity will weaken the amplitude of chiral condensate at certain temperature but slow its melting along the temperature.

{\it{Vector field under rotation and B field}.---}
In IMC, although the chiral condensate are highly non-perturbative, it could be qualitatively understood by noticing that light quark loops in the soft gluon polarization function would also be polarized by the magnetic field and quantitatively confirmed by the functional renormalization group computation\cite{Mueller:2015fka, Andersen:2021lnk} non-perturbatively. With effective models, such as NJL model, it has also been reproduced by introducing a magnetic dependent effective coupling constant. It is a more intuitive approach and the coupling running behavior has also been confirmed qualitatively by the straightforward computations of loop corrections. In the language of effective model the IMC is induced by the reduction of the coupling by background magnetic field. 
Explicitly the study with the NJL model shows the magnetic field will speed chiral condensate vanishing along temperature but also enhance the amplitude of condensation at a given temperature\cite{Farias:2014eca}.
%At given temperature the background magnetic field will enhance the chiral condensate, while along the temperature axis the condensate will melt more easily at larger magnetic field.    

For simplicity we will follow the strategy of effective model which requires the running behavior of the coupling constant as a function of rotation firstly. As mentioned above, we will extract it in a pure gluon system in this work. The equations of motion(EOM) of the vector particle under rotation  has been studied in \cite{Dai:2012bc,Huang:2020kik,Chernodub:2018era,jiangqed}. In the rotating frame the Lagrangian density should be the same as the free one by modifying the usual derivative to the covariant one which includes extra terms of the rotational connections. The EOM of the massive vector field reads
\begin{eqnarray}
&&\partial_i f_{i 0}-m^2 A_0=\Delta A_0-m^2 A_0=0
\end{eqnarray}
which gives $A_0=0$ for the transverse polarization components $A_{TE}$ and $A_{TM}$.
And 
\begin{eqnarray}
&&\partial^2_0 A_i-\Delta A_i -2v_j \partial_0\partial_j A_i+(\partial_j v_i-\partial_i v_j)\partial_0 A_j\nonumber\\
\label{roteq}
&&+v_j\partial_j(v_n\partial_n A_i+2\epsilon_{i n m}\omega_m A_n)-(\omega^2 A_i-\omega_n \omega_i A_n)\nonumber\\
&&+m^2 A_i=0
\end{eqnarray}
for the $\vec{A}$ part, where $f_{\mu\nu}=\partial_\mu A_\nu-\partial_\nu A_\mu$ 
and $\vec{v}=\vec{\omega}\times\vec{r}$ with $\vec\omega$ the angular velocity vector and $m$ the mass of the vector field. 
By considering the uniform rotation around the z-axis $\vec{\omega} =\omega \hat{e}_z$,  the EOM indicates in cylindrical coordinates $\vec{A}=A_\rho \hat{e}_\rho+A_\phi \hat{e}_\phi+A_z \hat{e}_z$ should satisfy
\begin{eqnarray}
(-E^2+\Omega^2 + m^2-2 E \omega n-{\omega}^2 n^2)A_{\rho, \phi, z}=0
\end{eqnarray}
where we have already assumed the angle and z-axis dependence of the eigen solutions are separated as 
$A_{\rho, \phi, z}=e^{i n \phi}e^{i k_z z}a_{\rho, \phi, z}(\rho)$.
The $\Omega$ is the eigen value of the $\Delta=\nabla^2$ operator, i.e. $-\Delta \vec{A}=\Omega^2\vec{A}$. Obviously the $\omega$ and $\omega^2$ terms are rotation relevant ones. The eigen energies could be read as $E=\pm\sqrt(\Omega^2+m^2)-n\omega$. This equation shows that no matter what the radial solutions in static($\omega=0$) are like, the rotation terms will not change the eigen states but just a shift of energies by $n\omega$ if the angle dependence of the solutions are chosen as $e^{i n \phi}$ in cylindrical coordinates. In \cite{jiangqed} we have already proven that when $m=0$ and $\omega=0$ the propagator of the vector field could be reduced to the usual gauge field propagator by discarding the $m^{-\alpha}$ terms and the longitudinal polarization component $A_L$. In the following we will set $m=0$ and focus on the $\vec{A}$ part to extract the eigen energies of the gauge field.

We further introduce the constant background magnetic field $\vec{B}=B\hat{e}_z$ as \cite{Nielsen:1980sx} which has no coupling with the rotation. The solutions are well-known Laudau levels in the asymmetric background magnetic field case, i.e. $\vec{A}=(0, Bx, 0)$. However we should work out eigen solutions in the symmetric scheme($\vec{A}=\frac{1}{2}(-By, Bx, 0)$) which could permit us to  choose the angle dependence of eigen functions as $e^{i n \phi}$ in cylindrical coordinates. The explicit form of the equation is 
\begin{eqnarray}
&&\left[\partial_\rho^2+\frac{1}{\rho}\partial_\rho-\frac{g^2 B^2}{4}\rho^2-\frac{(n\pm 1)^2}{\rho^2}+(n gB \pm 3 gB)\right]A_{\pm}\nonumber\\
&&=\left[-(E+n\omega)^2+k_z^2\right]A_{\pm}
\end{eqnarray}
where $A_\pm=A_\rho\pm i A_\phi$ and the $\pm$ terms corresponding to the two spin components of gauge field. This solution has also been studied in \cite{Chen:2015hfc} for a fermionic system under rotation and the real magnetic background field. Here as the original idea \cite{Nielsen:1980sx} suggests, the introduced magnetic field is actually the background color field $A_3^\mu$ for the $\text {SU}_c(2)$ gauge field and $A_8^\mu$ for the $\text{SU}_c(3)$ case. That is why the semi-classical background field has interaction with the dynamic gauge fields which carry no electric charges. The solutions are the Laguerre functions $L_{l\pm 1}^n(g B\rho^2/2)$ or the general confluent hypergeometric functions and the eigen energies are labelled by the $l\pm 1=\frac{E_n^2-m^2-k_z^2}{2 g B}$ where $E_n=E+n\omega$. 
In static case($\omega=0$) the system transverse size could be infinite and the boundary condition $A(r\rightarrow +\infty)=0$ requires the solutions to be Laguerre polynomials and thus $l$ non-negative integers. As the result eigen energies become the well-known Landau levels $E=\sqrt{2gB(l\pm 1)+k_z^2}$. 

Obviously in rotating system the rotational polarization shift of Landau levels $E=\sqrt{2gB(l\pm 1)+k_z^2}-n\omega$ have no lower bound. It is reasonable because the system should be strictly bounded by the light speed limit if the rotation is uniform. Otherwise if the system size is too large to the background rotation, one would expect it corrupt into several small ones with the same or lower angular velocity, which is known as lattice of vorticity. However the boundless will not bring serious issues to the chiral condensate computation because of the Bessel functions' asymptotic behaviors in finite temperature cases. For the zero temperature case the boundary will screen all the rotational effects on chiral condensate which is known as no rotation of the vacuum. While for the summation over energy levels it would be quite problematic. Therefore we should consider the boundary in this work.
In the uniformly rotating system we set the largest radius as $\rho=R$ whose corresponding largest vorticity is $\omega^*=R^{-1}$ because of the light speed limit. At the boundary the wave function should be zero $L_{l\pm 1}^n(g B R^2/2)=0$. Thus at given $n$, the $l$ should be chosen to make the $gB R^2/2$ locate at one of the zeros of Laguerre function $L_{l\pm 1}^n(z)$. This means the $l$ is not necessary integer but a implicit function of $n$ and $R$. However the eigen energies are the same form as the usual Landau levels $E=\sqrt{2gB(l(n, R)\pm 1)+k_z^2}-n\omega$ with $\omega<R^{-1}$. Numerically we will find the energies are always positive at any $n$. 

\begin{figure}[!hbt]
    \centering
    \includegraphics[width=7 cm]{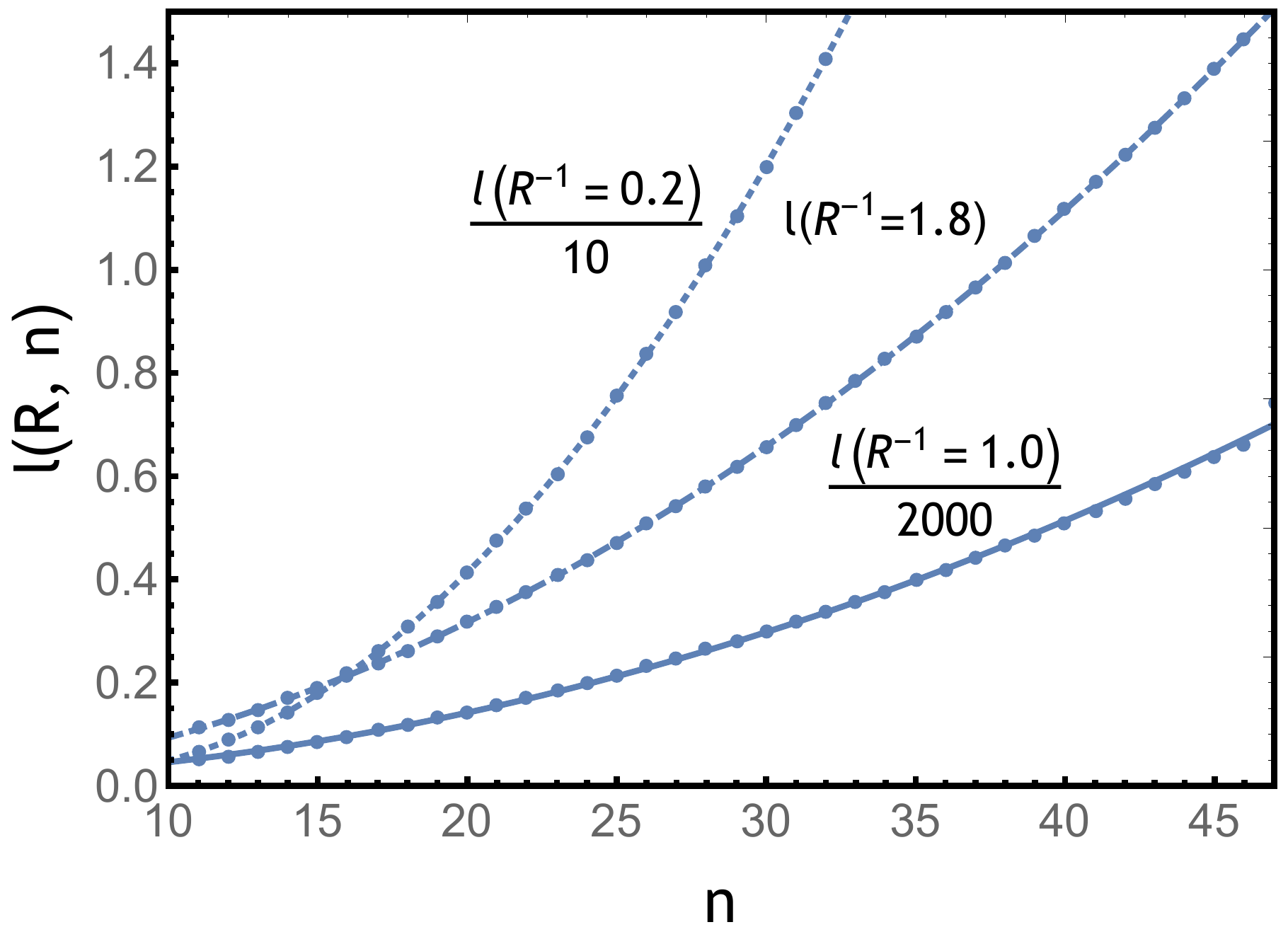}
    \caption{Fitting the $l(R=\{0.2^{-1},1.0^{-1},1.8^{-1}\}, n)$ with $l=a(R)+b(R)n+c(R)n^2$.}
    \label{fig: fitting1}
\end{figure}

\begin{figure}[!hbt]
    \centering
    \includegraphics[width=7 cm]{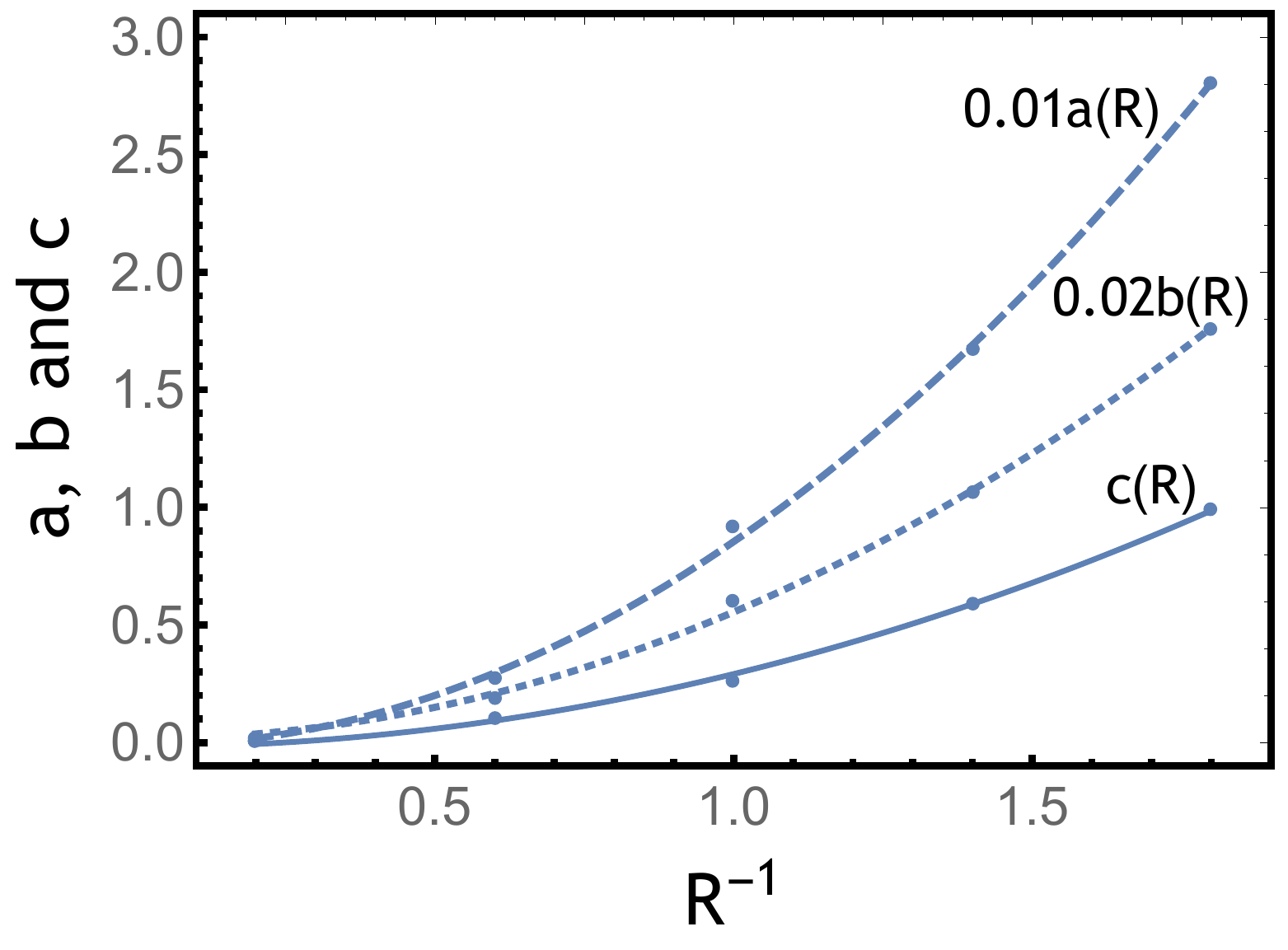}
    \caption{Fitting coefficients with $a(R)$, $b(R)$ and $c(R)$ with $R^{-2}$.}
    \label{fig: fitting2}
\end{figure}

To our knowledge there is no analytical results of $l$ for equation $L_{l\pm 1}^n(g B R^2/2)=0$. In order to obtain the vacuum energy in present of semi-classical color magnetic field, we choose the lowest level for each $n$ at different sizes $R$ and fit them to obtain a analytical expression. The numerical solutions(dots) and fitting curves are shown in Fig. \ref{fig: fitting1} and \ref{fig: fitting2}. It is clear that the fitting works very well because the dots of numerical solutions almost locate on the fitting curves precisely. The fitting result is
\begin{eqnarray}
\label{levels}
l(R, n)=\frac{1}{gB R^2}(87+15.5 n+0.54 n^2)
\end{eqnarray}
Noting that at large $n$ the Landau-level-like part $\sqrt{2gB(l(n, R)\pm 1)+k_z^2}\ge \sqrt{1.08}n R^{-1}$ which is always larger than the rotational polarization part $n\omega$ 
where $\omega\le R^{-1}$ because of light speed limit. This preserves the energy levels are always lower bounded with finite transverse size. 

{\it{Rotation dependence of the running coupling}.---}
In principle the coupling running is computed straightforwardly by the one-loop correction of the $\bar{\psi}\psi A$ vertex. However propagators in the cylindrical coordinates no longer have translation invariance which would have been translated to energy-momentum conservation at every vertex. In this case it is difficult to complete the loop integrals or summations analytically and further extract the $\gamma^\mu$ coefficient as the vertex correction because of lack of good summations results of multi-Bessel functions' products involved in the computation. Alternatively as a preliminary study for a qualitative result we follow the approach in \cite{Nielsen:1980sx} to extract the rotation dependence of the coupling which involves only the energy levels of single particle. Adopting the fitting result in Eq.\ref{levels} and including its density of state $\frac{1}{gB R^2}(15.5+1.08 n)$ we completed the summation over energy levels to yield the vacuum energy density in a background color field and under rotation as 
\begin{eqnarray}
E_{vac}=&&\frac{1}{8\pi^2}\int dk_z\sum_n \frac{15.5+1.08 n}{gB R^2}\theta(\Lambda-E)\nonumber\\
&&\times(\sqrt{2gB(l(n, R)\pm 1)+k_z^2}-n\omega)
\end{eqnarray}
where the $\Lambda$ is the ultra-violate cutoff. The longitudinal momentum $k_z$ can be integrated out analytically. And we further used the Euler summation formula to complete the summation over $n$. Obviously the summation is divergent which will also emerge in loop computations. We follow the same renormalization procedure in \cite{Nielsen:1980sx}to remove the $\Lambda$ and large $n$ divergent terms. Up to the $R^{-2}$ terms we obtain the energy density which is supposed to equal the energy of magnetization as
\begin{eqnarray}
E_{vac}&&=-\frac{(gB)^2}{8\pi^2}[ln(R\Lambda)+\frac{1.08}{12 gB R^2}\frac{\omega}{\Lambda}]\nonumber\\
&&=-\frac{1}{2}4\pi\chi B^2
\end{eqnarray}
where $\chi$ is the effective magnetic susceptibility. Hence the effective dielectric constant is obtained
\begin{eqnarray}
\epsilon=\frac{1}{1+4\pi\chi}=[1+\frac{g^2}{4\pi^2}(ln(R\Lambda)+\frac{1.08}{12 gB R^2}\frac{\omega}{\Lambda})]^{-1}
\end{eqnarray}
By comparing coefficients the rotation dependence of the running coupling is
\begin{eqnarray}
\label{running}
\alpha_{eff}=\alpha[1+\frac{\alpha}{\pi}(ln(R\Lambda)+\frac{1.08}{6\langle k^2\rangle R^2}\frac{\omega}{\Lambda})]
\end{eqnarray}
Here we have replaced the background field $2gB$ with the average momentum scale of the system $\langle k^2\rangle$ as the standard procedure.
It shows in fixed transverse size system it increases with angular velocity. And as a byproduct it is clear that in static case the coupling decreases with $R$, which is consistent with the famous asymptotic freedom. 
In \cite{Schneider:2002jc,Schneider:2002sg,Schneider:2003uz} this method has been extended to finite temperature and density systems. Although in the following the chiral condensate behavior will be studied at
finite temperature, we only take the qualitatively rotation dependence in Eq.\ref{running} for a preliminary study and explore the existence of a non-trivial impact on the
chiral transition. Technically more detailed studies could be done following works in \cite{Schneider:2002jc,Schneider:2002sg,Schneider:2003uz}.
In the following we consider a strong interacting system of many chiral fermions at finite temperature to choose $\langle k^2\rangle R^2\sim 1$. With effective model we will study the pseduo critical behavior of chiral condensate as a function of temperature and rotation. 

{\it{Chiral rotational catalysis in NJL model}.---}
Comparing with the IMC results, the coupling increasing with rotation suggests that the chiral condensate would melt more slowly with temperature. Equivalently it is expected that the critical temperature of chiral restoration will also increase with rotation. Comparing to the IMC we name this behavior as chiral vortical catalysis. In order to show this numerically we choose the NJL model with traditional parameters $\Lambda=0.65$GeV, $m=0.005$GeV and set the coupling running as
\begin{eqnarray}
G(\omega)=G(1+0.32\frac{\omega}{\Lambda_{NJL}})
\end{eqnarray}
where $G=4.93\text{GeV}^{-2}$.
Adopting the usual mean field approximation $M=m-2G\langle\bar\psi \psi\rangle$ and local potential approximation(LPA), the gap equation for the chiral condensate has already been studied in \cite{Jiang:2016wvv}. We write it down directly as
\begin{eqnarray}
M=m_0-4GM\sum_{m k_z k_s}&&\frac{2n_f(E_{k m})-1}{E_k}\nonumber\\
&&\times(J^2_m(k_t r)+J^2_{m+1}(k_t r))
\end{eqnarray}
where $\sum_{n k_z k_s}=\frac{1}{8\pi^2}\int k_t dk_t dk_z\sum_{n=-\infty}^\infty$ and $E_{kn}=\sqrt{k_t^2+k_z^2+m^2}-(n+\frac{1}{2})\omega$. Going beyond LPA the self-consistent position dependence of the condensate can be studied as \cite{Wang:2018zrn} which indicates the LPA's results qualitatively agree with the self-consistent ones locally. Here we also consider the finite size case the integral over $k_t$ should be replaced by summation over the zeros of $J_m(k_t R)=0$. And we choose $R=2GeV^{-1}$ and compute the condensate at $r=0.1GeV^{-1}$.

\begin{figure}[!hbt]
    \centering
    \includegraphics[width=7 cm]{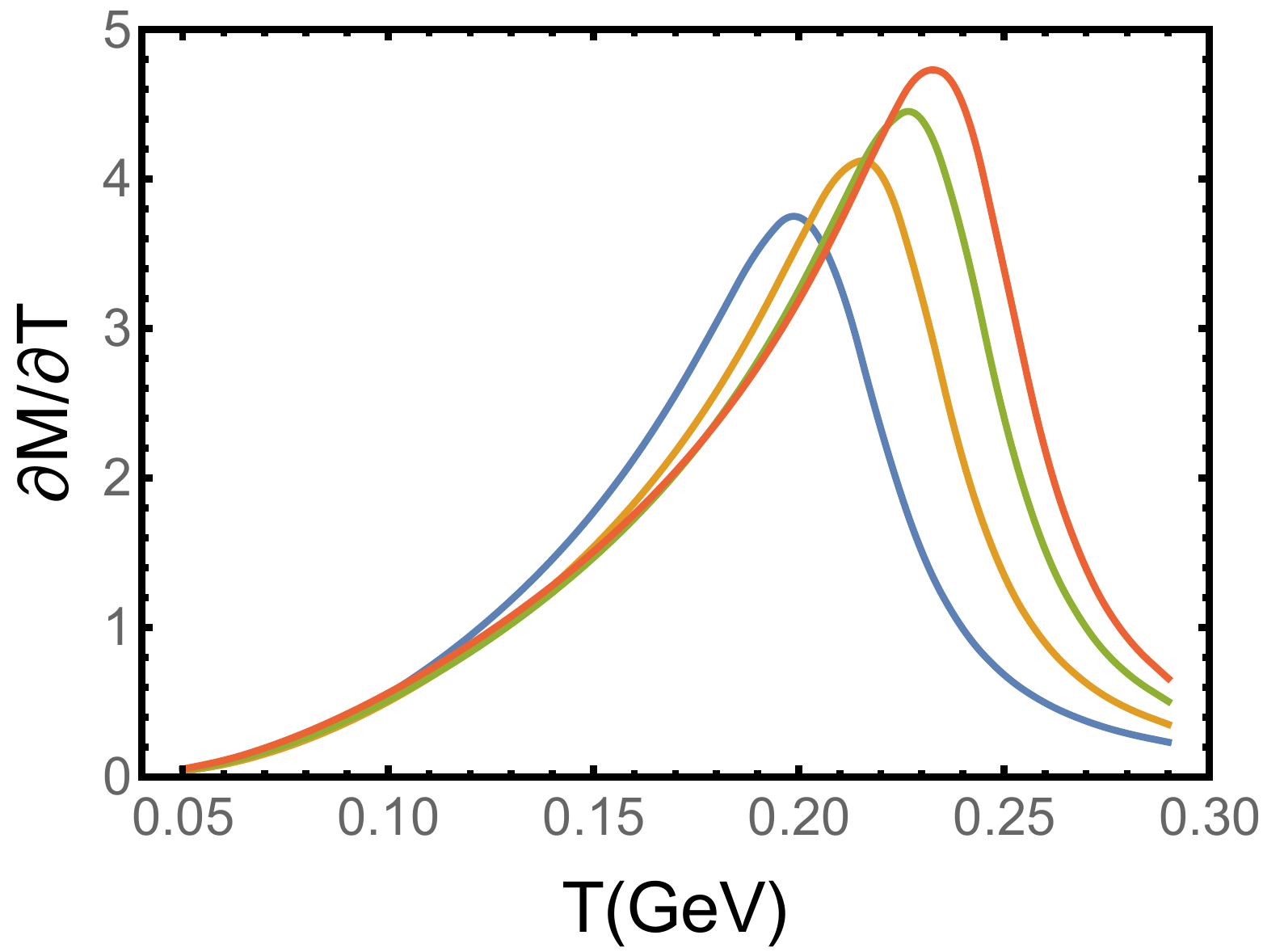}
    \caption{Chiral susceptibility as functions of temperature at  $\omega=0.1, 0.2, 0.3, 0.4$GeV(from left to right).}
    \label{fig: chi}
\end{figure}

\begin{figure}[!hbt]
    \centering
    \includegraphics[width=7 cm]{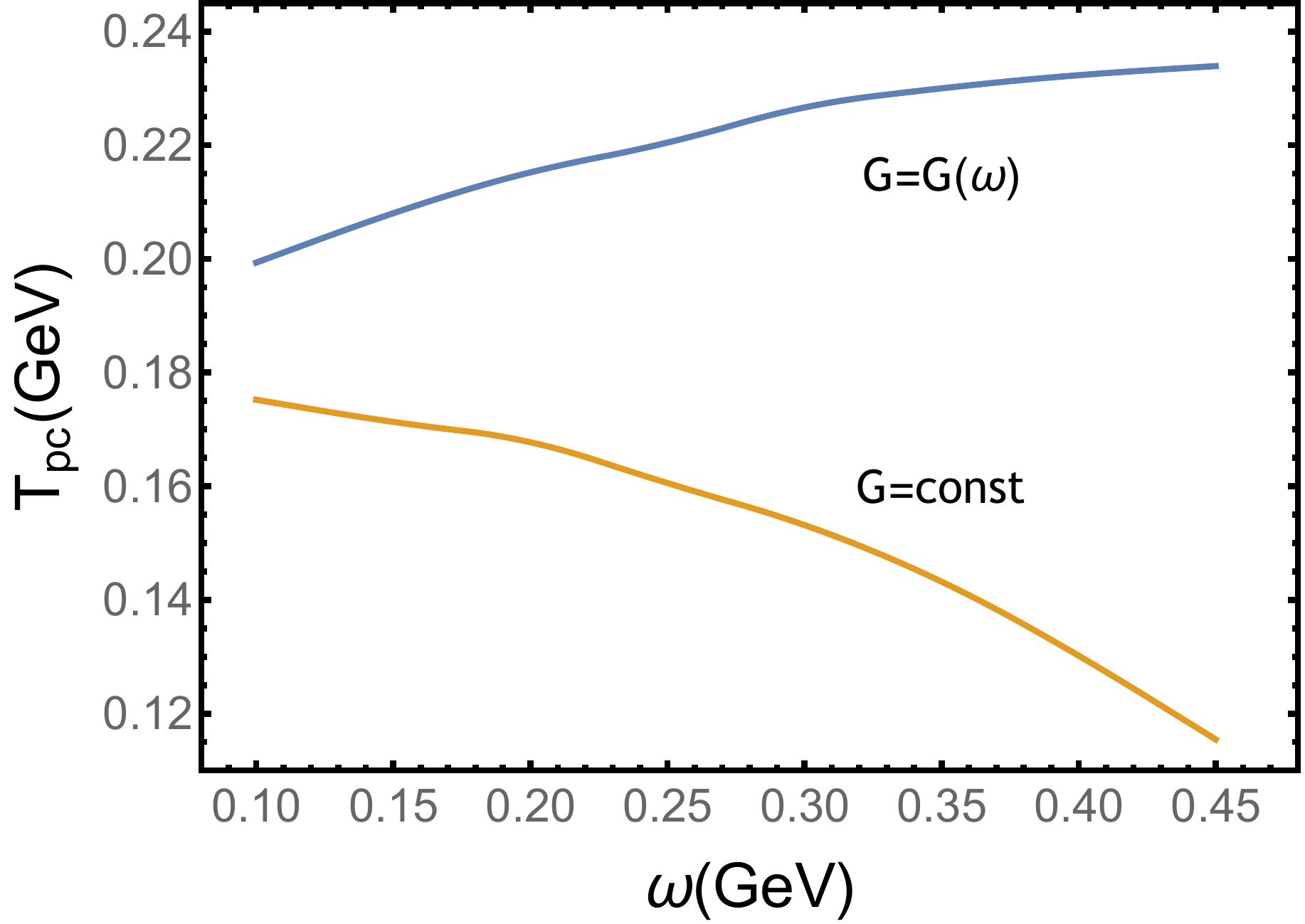}
    \caption{Pseudo critical temperature as function of $\omega$.}
    \label{fig: tcp}
\end{figure}

We plot both the chiral susceptibility(Fig.\ref{fig: chi}) and the corresponding pseudo critical temperatures(Fig.\ref{fig: tcp}) as follows. Because of a approximate chiral symmetry for the small current mass value the pseduo critical temperature is define as the peak location of chiral susceptibility. As expected, from the Fig.\ref{fig: chi} one could find that the peak moves towards right with faster rotation. By plotting the pseudo critical temperature as the function of rotation the Fig.\ref{fig: tcp} shows the trend of catalysis more clearly and eventually appears to be saturated while in the constant coupling case decreasing monotonously. However one should treat the saturation carefully because of the rotation approaching the cutoff there. 

{\it{Summary and Discussion}}
Using the semi-classical background field method we studied the rotation dependence of the effective strong interaction coupling. It is shown that  the effective coupling will become larger with the rotation increasing which is contrary to the magnetic field case. As the physical result the increasing coupling will slow the condensate melting procedure along temperature. This could be named as the chiral vortical catalysis or inverse rotational suppression. Imposing the running behavior to the coupling of the NJL model, we numerically computed the chiral condensate and showed this catalysis scenario in the intuitive mean field approximation. The pseudo critical temperature is confirmed to rise with rotation and approaches a saturation eventually which may be induced by the model cutoff.

%{\color{red}For vorticity itself, after a inspiring qualitative agreement of the simulations of the $\Lambda$ polarization dependence on the collision energy and centrality with the STAR measurements\cite{STAR:2017ckg,Xie:2017upb}, people found that the vorticity distribution and evolution in QGP may be much more complicated by considering both the quantitative analysis on the global polarization and the qualitative mismatch of the local polarization profiles\cite{Shapoval:2017jej,Li:2017dan, Becattini:2020ngo}.}
%In order to understand the vector meson polarization measurements\cite{BedangadasMohantyfortheALICE:2017xgh,Zhou:2019lun,Mohanty:2020bqq}, in the framework of kinetic theory, quantum Wigner functions and hydrodynamics, various attempts\cite{Kharzeev:2013ffa,Li:2017dan,Karpenko:2021wdm, Avkhadiev:2017fxj, Gao:2019znl,Weickgenannt:2019dks,DelZanna:2013eua,Huang:2018aly,Huang:2020kik, Karpenko:2013wva} have been made to mend the usual simulation which is basing on the straightforward scenario of the parton collisions. In most of these works, valence quraks, which serve as spin-carriers of final-state hadrons, have attracted almost all the interests\cite{Yang:2017sdk,Sheng:2019kmk}. The gluons, which carrying double spin number and thus suffered double polarization effects, are not treated seriously enough. 

In the context of QCD strong interaction coupling is crucial in different systmes, from the phase structure in thermalized system to the hadron production during the heavy ion collision. The rotation dependence study should be further extended to different number of colors  and light quark flavors as well as different thermal conditions, such as finite temperature, density and magnetic field cases. One potential measurable signal could be the vector $\phi$ and $K^{*0}$ polarization in heavy ion collisions\cite{BedangadasMohantyfortheALICE:2017xgh,Zhou:2019lun,Mohanty:2020bqq,Shapoval:2017jej,Sheng:2019kmk}. Considering this surprising polarization measurements of vector hadrons with different strangeness number, the rotation and the flavor dependence of coupling may play an important role in the hadronization process.  We will extend the study to these problems with both this framework and other systematic non-perturbative approaches, such as functional renormalization group, in future. Apart from the high energy topics, the non-abelian gauge interaction has also been realized in the cold atom and condensed matter systems. We expect the non-trivial coupling running behavior would be more interesting in such system both theoretically and experimentally because of much better maneuverability of them. Finally one should note that the rotating QCD system has no sign problem although the rotation relevant terms appear as a effective chemical potential in the dispersion relation. This means it could be simulated and checked with lattice QCD as the magnetic case.  

{\bf Acknowledgments.}
The work of this research is supported by the National Natural Science Foundation of China, Grant Nos. 11875002(YJ) and the Zhuobai Program of Beihang University.

\end{document}